\documentclass[oneside]{amsart}

\usepackage{amssymb}

\theoremstyle{plain}    
\newtheorem{thm}{Theorem}[section]
\newtheorem*{fact*}{Fact} 
\newtheorem{lem}[thm]{Lemma}
\newtheorem*{cor*}{Corollary}
\newtheorem{cor}[thm]{Corollary}
\newtheorem{example}[thm]{Example}

\theoremstyle{definition}
\newtheorem*{defn*}{Definition}
\newtheorem{defn}[thm]{Definition}

\numberwithin{equation}{section}
\numberwithin{figure}{section}

\newcommand{\A}{\mathcal{A}}

\newcommand{\SU}{\textup{SU}}
\newcommand{\SO}{\textup{SO}}
\newcommand{\Spin}{\textup{Spin}}
\newcommand{\tensor}{\otimes}
\newcommand{\ft}{\mathcal{F}}
\newcommand{\ftinv}{\ft^{-1}}
\newcommand{\R}{\mathbb{R}}
\newcommand{\lir}{L^1(\R)}
\newcommand{\ltr}{L^2(\R)}
\newcommand{\lpr}{L^p(\R)}
\newcommand{\ra}{\rightarrow}

\newcommand{\intii}{\int_{\!-\infty}^{\infty}\!}
\newcommand{\intzi}{\int_0^{\infty}\!}
\newcommand{\prodin}{\prod_{i=1}^n}
\renewcommand{\d}{\,d}
\newcommand{\dy}{\d y}
\newcommand{\dx}{\d x}
\newcommand{\dt}{\d t}
\newcommand{\dk}{\d k}
\newcommand{\dl}{\d \ell}
\newcommand{\dr}{\d r}
\newcommand{\ddk}{\frac{d}{dk}}
\newcommand{\dddk}{\frac{d^2}{dk^2}}
\newcommand{\pv}{\textup{pv}}
\newcommand{\pvk}{\pv\frac{1}{k}}
\newcommand{\pvik}{\pv\frac{i}{k}}

\newcommand{\pvintii}{\pv\intii}
\newcommand{\delk}{\delta(k)}
\newcommand{\hsmash}[1]{\makebox[0pt][l]{#1}}
\newcommand{\pf}{p_{\mspace{-2mu}f}}
\newcommand{\dpf}{\d\pf}
\newcommand{\limeps}{\lim_{\epsilon\ra0\hsmash{$^{\!+}$}}\ }
\newcommand{\intieei}{\left( \int_{\!-\infty}^{-\epsilon} + \int_{\epsilon}^{\infty} \right)}
\newcommand{\intAeeA}{\left( \int_{-A}^{-\epsilon} + \int_{\epsilon}^{A} \right)}
\newcommand{\feps}{F_\epsilon}
\newcommand{\HH}{\mathbb H}
\newcommand{\Ht}{{\HH^3}}
\newcommand{\intH}{\int_\Ht}

\newcommand{\dfn}[1]{\textbf{\mathversion{bold}#1}}

\DeclareMathOperator{\conv}{\ast}
\DeclareMathOperator{\convin}{\overset{n}{\underset{i=1}\ast}}
\DeclareMathOperator{\sinc}{sinc}

\newcommand{\longpage}{\enlargethispage{\baselineskip}}

\DeclareMathOperator*{\fprod}{\overline{\prod}}

\begin{document}

\title{Positivity in Lorentzian Barrett-Crane Models of Quantum Gravity}

\author{J. Wade Cherrington}
\address{Department of Mathematics \\
University of Western Ontario \\
London, Ontario, Canada }
\email{jcherrin@uwo.ca}

\author{J. Daniel Christensen}
\address{Department of Mathematics \\
University of Western Ontario \\
London, Ontario, Canada }
\email{jdc@uwo.ca}

\begin{abstract}
The Barrett-Crane models of Lorentzian quantum gravity are a family
of spin foam models based on the Lorentz group.
We show that for various choices of edge and face amplitudes, 
including the Perez-Rovelli normalization, the amplitude for every
triangulated closed $4$-manifold is a non-negative real number.
Roughly speaking, this means that if one sums over triangulations,
there is no interference between the different triangulations.
We prove non-negativity by transforming the model into a ``dual variables'' 
formulation in which the amplitude for a given triangulation is 
expressed as an integral over three copies of hyperbolic space for each 
tetrahedron.
Then we prove that, expressed in this way, the integrand is non-negative. 
In addition to implying that the amplitude is non-negative, 
the non-negativity of the integrand is highly significant 
from the point of view of numerical computations, as
it allows statistical methods such as the Metropolis algorithm
to be used for efficient computation of expectation values of
observables.
\end{abstract}

\maketitle

\section{Introduction}

Spin foam models provide a background independent approach to quantum
field theory and arise naturally as a path integral formulation of
quantum gravity~\cite{Baez,Baez2,BC,BC2,PR,RR}.  

This paper deals with the Lorentzian Barrett-Crane model, but to set
the stage we first give some background on the Riemannian model.
The Riemannian model is not expected to be physically realistic, but has
been an important arena for testing theoretical and computational
aspects of the spin foam program.
Shortly after Barrett and Crane proposed their Riemannian model in~\cite{BC}, 
two specific choices of normalization factors for the amplitude were 
given in~\cite{DFKR} and~\cite{PR}; we call these the DFKR normalization
and the Perez-Rovelli normalization, respectively.
The DFKR normalization is believed to lead to divergent
amplitudes $\A(\Delta)$ for most triangulations $\Delta$;
this has been tested numerically~\cite{BaezChristensen}
and is thought to reflect some residual gauge symmetry~\cite{FL}.
The Perez-Rovelli normalization has been proven~\cite{P} to give
a finite amplitude $\A(\Delta)$ for each triangulation, but
it is not known how to regulate the sum over all triangulations.
For both normalizations, it has been shown~\cite{BCpos} that there 
is \emph{no interference} in the path integral.  

In order to elaborate on the last point, we recall some background.
Barrett-Crane spin foam models are generally
expressed as a sum over certain triangulations $\Delta$ of an amplitude
$\A(\Delta)$, and the amplitude $\A(\Delta)$ is itself a sum (or
integral) over labellings of faces of $\Delta$ by 
representations of a group.
In the sum, one restricts to triangulations of 4-manifolds
which interpolate between chosen initial and final 3-geometries.
In the case of the Riemannian model, the group is 
$\Spin(4) = \SU(2) \times \SU(2)$, which is the double-cover of
$\SO(4)$, and we restrict to the balanced representations 
$j \tensor j$ which are indexed by spins $j$.
Thus the amplitude can be expressed as
\[
  \A \equiv \sum_{\Delta} \A(\Delta) ,
\]
where
\begin{equation}\label{eq:sumjF}
  \A(\Delta) \equiv \sum_{j_f} \A(\Delta, j_f) .
\end{equation}
Here $j_f$ is a labelling of the triangles of $\Delta$ with spins,
and the details of $\A(\Delta, j_f)$ depend on the specific
normalization chosen.

What was shown in~\cite{BCpos} is that when the quantity 
$\A(\Delta, j_f)$ is nonzero, its sign depends only on the 
initial and final geometries (with their spin labellings), 
and not on the interpolating manifold, its triangulation
$\Delta$, or its spin labelling $j_f$.  This is the very strong
sense in which there is no interference in the Riemannian 
Barrett-Crane path integrals.  The conceptual meaning of this
result is not fully understood, especially since the model
is based on a real-time ($e^{iS}$) path integral
rather than an imaginary-time ($e^{-S}$) path integral.
Computationally, the result is extremely powerful, allowing one to use
statistical methods such as the Metropolis algorithm for calculating
expectation values~\cite{BaezChristensen}.

Now we describe the current state of knowledge of the Lorentzian
Barrett-Crane model.
This model was defined by Barrett and Crane in~\cite{BC2}
and involves representations of the connected Lorentz
group $\SO_0(3,1)$ from the principal series.  These
representations are indexed by non-negative real numbers $p$,
and so the sum in equation~(\ref{eq:sumjF}) becomes an
integral over a product of copies of $\R^+$.
In~\cite{perez-rovelli}, a specific normalization was
given, which we again call the Perez-Rovelli normalization.
With this normalization, the Lorentzian Barrett-Crane
model was proved~\cite{CPR,CPRpub} to give
a finite amplitude $\A(\Delta)$ for each triangulation.
In~\cite{wade} this proof was extended to include certain
degenerate triangulations.
As in the Riemannian case,
it is not known how to regulate the sum over triangulations.

In~\cite{BCpos} numerical computations were done
which give evidence to the conjecture that there is no
interference in the Lorentzian path integral, i.e., that
when the quantities $\A(\Delta, \pf)$ are nonzero, their
sign depends only on the initial and final geometries.
In fact, these amplitudes seem to always be non-negative.
However, this is still an open question.

In the present paper, we focus on the case of closed 4-manifolds
and show that after transforming to a ``dual variables'' formulation, 
the model then has the property that
there is no interference in the new path integral. 
We also have partial results in the case of manifolds with boundary.

The dual variables formulation, introduced in~\cite{wade}, is 
essentially a specific realization (for Lorentzian spin foams) 
of a transformation first proposed by Pfeiffer~\cite{pfeiffer} in the
context of Riemannian spin foams.  
In the dual variables formulation, $\A(\Delta)$ is
expressed as an integral with respect to hyperboloid variables as follows.  
For each tetrahedron $e$ of $\Delta$, there is an associated set of
three variables $x_e^i=(x_e^0, x_e^1, x_e^2)$.  Each of these
variables takes values on the future 3-hyperboloid.
The first comes from the edge amplitude, and the other two are
associated to the two 4-simplices that $e$ is contained in, and
come from the vertex amplitude.

The transformation to dual variables allows us to define an amplitude $\A(\Delta, x_e^i)$
such that $\A(\Delta)$ is the integral of $\A(\Delta, x_e^i)$
with respect to the $x_e^i$ variables.
In this paper, we prove that the quantities $\A(\Delta, x_e^i)$ are always non-negative.
While this doesn't tell us anything about interference
between the amplitudes $\A(\Delta, \pf)$, it does
imply that in the original path integral there is no interference
between different triangulations.
That is, we find that the quantities $\A(\Delta)$ are always
non-negative.

As we have explained, our result is not exactly analogous to the result 
in the Riemannian case, in that we have not shown that there is no
interference between the amplitudes $\A(\Delta, \pf)$ 
(although we still believe this to be true) but only that 
there is no interference between the amplitudes $\A(\Delta, x_e^i)$.
At first, this would seem to not be enough to allow statistical 
methods to be used for the computation of expectation values of
observables, especially observables which are naturally functions
of the $\pf$ variables.
But it turns out that in many cases the computation of such an 
expectation value can also be transformed into the dual variables 
formulation, where there is no interference, and that statistical
methods can be successfully applied.
Using this, the first author has made the first ever 
computations of expectation values of geometric observables in a 
Lorentzian Barrett-Crane model.
These will be the subject of a forthcoming paper~\cite{wade2}.

In addition to proving the above results for the Perez-Rovelli model,
we illustrate the robustness of the methods by handling models
with the non-standard face normalizations $\A_f=1/\pf$ and $\A_f=1/\pf^2$.
The methods could be used to check for positivity in other normalizations 
as well.

Next we outline the paper.  
In Section~\ref{se:ffact} we recall the Lorentzian Barrett-Crane
model, describe the transformation to dual variables,
and show how the results of Section~\ref{se:pos} imply 
our positivity results.
In Section~\ref{se:ft} we give some elementary results
about Fourier transforms that we will need.
In Section~\ref{se:pos} we prove several positivity results,
with each proof corresponding to a different normalization.
We give conclusions in Section~\ref{se:concl} and prove some
technical results about Fourier transforms in 
Appendices~\ref{ap:evenbumps} and~\ref{ap:pv}.

\section{Positivity in the Dual Variables Form of the Perez-Rovelli Model}
\label{se:ffact}

In this section, we review the Perez-Rovelli model and its dual formulation
in terms of hyperboloid variables~\cite{wade}.
Then we show how the results of Section~\ref{se:pos} can be
used to show that there is no interference in the dual path integral.
As we shall see, in the case of a manifold with boundary,
the conclusion depends on whether the boundary
data is in terms of hyperboloid labellings or representation labellings.

For simplicity, we assume throughout that our triangulations are
non-degenerate, but this can be relaxed in many cases.

\subsection{The Perez-Rovelli Model}

We start by recalling the general form of the Lorentzian Barrett-Crane model,
which assigns to each triangulation $\Delta$ of a 4-manifold an amplitude.
For now we deal with closed 4-manifolds; we discuss manifolds with
boundary in Subsection~\ref{ss:boundary}.
To be consistent with cited papers, we work with the dual 2-skeleton
of the triangulation $\Delta$:  to each 4-simplex, we associate a
dual vertex; to each tetrahedron, we associate a dual edge; and to
each triangle, we associate a dual (polygonal) face.  
$V$, $E$ and $F$ denote the sets of dual vertices, dual edges and
dual faces, respectively.
(This use of the word ``dual'' has nothing to do with the phrase 
``dual variables''.)

The general Lorentzian Barrett-Crane amplitude is
\begin{equation} \label{eq:BCstatesum}
  \A(\Delta)\equiv\underbrace{\intzi\cdots\intzi}_{f\in F}
  \Biggl(\prod_{f\in F}\A_f\Biggr) \Biggl(\prod_{e\in E}\A_e\Biggr)
  \Biggl(\prod_{v\in V}\A_v\Biggr) \prod_{f\in F}\pf^2 \dpf,
\end{equation}
where the factors of the form $\pf^2$ arise from the measure
on the principal series of $\SO_0(3,1)$ representations and the amplitudes
$\A_f$, $\A_e$, and $\A_v$ will be defined shortly. 
It should be noted that while some authors absorb the $\pf^2$ factors
from the measure into the definition of $\A_f$,
in the present work we keep the measure and the face amplitude distinct.

To more compactly represent the multiple integrations that take place
in~(\ref{eq:BCstatesum}) and in the amplitudes, we introduce the symbol $\fprod$
to indicate a formal product of symbols such as integral signs or
measures. When a relation such as $f\ni v$ appears below a product
symbol (multiplicative or formal), the product is taken over all of
the objects on the left hand side of relation that satisfy the relation;
for example, $\prod_{f \ni v}$ denotes a product over all dual faces
$f$ such that $v$ is a member of $f$. 

We now turn to the definition of the amplitudes $\A_v$,
$\A_e$, and $\A_f$ that appear in~(\ref{eq:BCstatesum}). 
First we need the following notation.  
We denote hyperbolic space by
$\Ht \equiv \{ x \in \R^4 \mid x \cdot x = 1 \text{ and } x_0 > 0 \}$,
where $x\cdot y$ is the Minkowski inner product 
$x \cdot y \equiv x_0 y_0 - x_1 y_1 - x_2 y_2 - x_3 y_3$. 
If $x$ and $y$ are points in hyperbolic space, then
$\phi(x,y)$ denotes the hyperbolic distance $\cosh^{-1}(x \cdot y)$
between $x$ and $y$.
If $v$ is a dual vertex which is contained in the dual face $f$, 
then $\phi_v^f$ is $\phi(x_{e_1}, x_{e_2})$, 
where $e_1$ and $e_2$ are the two edges of the polygon $f$ which meet at $v$; 
equivalently, in the triangulation picture, $e_1$ and $e_2$ are the two 
tetrahedra in the 4-simplex $v$ which share a common triangle $f$.

Up to a regularization that will be defined shortly, the Lorentzian 
Barrett-Crane vertex amplitude of~\cite{BC2} is defined as
\begin{equation}\label{eq:vertexampl}
  \A_v(\pf)\equiv\Biggl(\fprod_{e \ni v} \intH \dx_e \Biggr)
            \prod_{f \ni v} K_{\pf}(\phi_v^f) ,
\end{equation}
where the kernel function $K_{\pf}$ is 
\begin{equation}\label{eq:kernel}
  K_{\pf}(\phi_v^f)\equiv\frac{\sin(\pf\,\phi_v^f)}{\pf\sinh(\phi_v^f)}.
\end{equation}
As there are ten triangles in the 4-simplex to which a vertex is dual, 
the integrand is a product of ten such kernels. 

Although the expression for the vertex amplitude given above is generally
infinite, we adopt the usual regularization by fixing the value of
one $\Ht$ variable and dropping the corresponding integral
over $\Ht$---the answer is independent of the choices~\cite{BC2}.

When dealing with several of these integrals at once, we distinguish
the dummy variables of integration $x_e$ contributed by different
dual vertices $v$ by denoting them $x_e^v$.

While the original model of~\cite{BC2} specifies the vertex
amplitude $\A_v$ as given in~(\ref{eq:vertexampl}), it leaves unspecified
the edge and face amplitudes $\A_e$ and $\A_f$.
The most common choices of these are due to Perez and Rovelli. 
In this model~\cite{perez-rovelli},
Perez and Rovelli specify $\A_f=1$ and $\A_e=\Theta_4(p_1,p_2,p_3,p_4)$,
where $p_1,\ldots,p_4$ are the representation variables labelling the four
faces containing $e$.
The function $\Theta_4(p_1,p_2,p_3,p_4)$
is known as the \emph{eye diagram} and can be defined as follows:
\begin{equation}\label{eq:eye}
\begin{split}
  \Theta_4(p_1,p_2,p_3,p_4)  
      &\equiv \frac{1}{2 \pi^2} \intH
         K_{p_1}(\phi(x,y)) K_{p_2}(\phi(x,y))
         K_{p_3}(\phi(x,y)) K_{p_4}(\phi(x,y)) \dx \\
      &= \frac{1}{2 \pi^2} \intH
         \frac{\sin(p_1\phi(x,y))\sin(p_2\phi(x,y))\sin(p_3\phi(x,y))\sin(p_4\phi(x,y))}
              {p_1 p_2 p_3 p_4 \sinh^4(\phi(x,y))} \dx \\
      &= \frac{2}{\pi} \intzi 
         \frac{\sin(p_1 r)\sin(p_2 r)\sin(p_3 r)\sin(p_4 r)}
              {p_1 p_2 p_3 p_4 \sinh^2(r)} \dr .
\end{split}
\end{equation}
The value is independent of the choice of $y \in \Ht$.
For simplicity, we take $y$ to be the origin $O$ of hyperbolic space, so
$r = \phi(x, O)$ is the radial coordinate of $x$.
When dealing with several of these integrals at once, we distinguish
the dummy variables of integration $x$ and $r$ by denoting them
$x_e^0$ and~$r_e$.

Inserting~(\ref{eq:vertexampl}), (\ref{eq:kernel}), $\A_f=1$ and~(\ref{eq:eye})
into~(\ref{eq:BCstatesum}) gives
\begin{multline}\label{eq:PRstatesum}
\A(\Delta) =   \Biggl(\fprod_{f\in F} \intzi \dpf \Biggr)
               \Biggl(\prod_{f\in F} \pf^2 \Biggr)
               \Biggl(\prod_{e\in E} 
                       \intzi \dr_e \frac{2}{\pi \sinh^2(r_e)} 
                          \Biggl(\prod_{f \ni e} \frac{\sin(p_f r_e)}{p_f} \Biggr) 
               \Biggr) \\
               \Biggl(\fprod_{v\in V} \Biggl(\fprod_{e \ni v, e \neq e_0^v}
                      \intH \dx_e^v \Biggr)
                      \Biggl(\prod_{f \ni v}K_{\pf}(\phi_v^f)\Biggr)\Biggr)
\end{multline}
As called for by the regularization, for each dual vertex $v$, 
a dual edge $e_0^v$ is chosen and the integral with respect to 
$x_{e_0^v}^v$ is omitted.

We shall henceforth refer to this choice of amplitudes as the \emph{Perez-Rovelli Model}. 

\subsection{Face Factoring and Positivity for Closed Manifolds}
\label{ss:ffact}

Equation~(\ref{eq:PRstatesum}) defines
the Barrett-Crane amplitude as the result of first integrating
with respect to the hyperboloid variables $x_e^0$ (or, equivalently, $r_e$)
and $x_e^v$, and then integrating
the result with respect to the representation variables $\pf$.
This order of integration has provided the context for most
theoretical and numerical work to date.
In~\cite{wade}, the integrals are shown to be absolutely
convergent for all non-degenerate 4-manifold triangulations,
and so it follows that the order of integration can be reversed.
That is, we can integrate with respect to the representation variables 
$\pf$ first, and regard the integration with respect to the 
variables $r_e$ and $x_e^v$ as the path integral.
The result is 
\begin{equation}\label{eq:facefactoring}
\A(\Delta) =  \Biggl(\fprod_e \intzi \dr_e\Biggr)
                \Biggl(\fprod_v
                  \Biggl(\fprod_{e\ni v,e\neq e_0^v}\intH\dx_e^v\Biggr)\Biggr)
                  \A(\Delta, x_e^i)
\end{equation}
where
\begin{equation}\label{eq:Axei}
\A(\Delta, x_e^i) \equiv \Biggl(\prod_e\frac{2}{\pi \sinh^2(r_e)}\Biggr)
              \Biggl(\prod_v
                  \Biggl(\prod_{f\ni v}\frac{1}{\sinh(\phi_v^f)}\Biggr)\Biggr)
                \Biggl(\prod_f\intzi F_f(\pf,\phi_v^f,r_e) \dpf\Biggr)
\end{equation}
and
\begin{equation}\label{eq:facefactor}
  F_f(\pf,\phi_v^f,r_e) \equiv 
  \frac{\sin(\pf\,\phi_{v(f,1)}^f)\cdots\sin(\pf\,\phi_{v(f,\deg_V(f))}^f)
  \sin(\pf\,r_{e(f,1)})\cdots\sin(\pf\,r_{e(f,\deg_E(f))})}
  {\pf^{\deg_V(f)+\deg_E(f)-2}}.
\end{equation}
We have factored the integrand into \emph{face factors}
$F_f(\pf,\phi_v^f,r_e)$ that each depend on only a single $\pf$ variable
(but depend on several $x_e^v$ and $r_e = \phi(x_e^0,O)$ variables).
In~(\ref{eq:facefactor}), $\deg_E(f)$ denotes the number of edges contained in the
face $f$ and $e(f,i)$ selects the $i$th edge contained in the
face $f$. Similarly,%
\footnote{For closed 4-manifolds, dual faces are closed polygons so $\deg_E(f)=\deg_V(f)$.
If the 4-manifold has a boundary, the dual object to a triangle in the
boundary is an ``open polygon'' with free edges. In this
case, $\deg_E(f)=\deg_V(f)+1$.}
$\deg_V(f)$ denotes the number of vertices
in the face $f$ and $v(f,i)$ selects the $i$th edge contained
in the face $f$.
The number of factors in the denominator is two less than in the
numerator because of the $\pf^2$ in the measure. 
The integrals $\intzi F_f(\pf,\phi_v^f,r_e) \dpf$,
which we shall refer to as \emph{integrated face factors,} can be
found exactly in closed form~\cite{wade}, which is particularly
useful for numerical applications.

As we shall see in Subsection~\ref{ss:1}, 
an analysis of the face factors in the Fourier domain 
shows that the integrated face factors 
$\intzi F_f(\pf,\phi_v^f,r_e) \dpf$ are non-negative
when $\deg_V(f) + \deg_E(f) \geq 3$, which is always the case
for non-degenerate triangulations.
It follows immediately that each amplitude $\A(\Delta, x_e^i)$ 
is non-negative, and therefore that the amplitude $\A(\Delta)$
for any triangulation of a closed 4-manifold is non-negative.
Physically, this means that there is no interference between 
different triangulations $\Delta$.
This result has not previously been shown, although there were 
strong indications from numerical work~\cite{BaezChristensen}
as pointed out in the introduction.

The non-negativity of $\A(\Delta,x_e^i)$ 
means that for a given triangulation $\Delta$, there is no interference
amongst any of the $x_e^i$ configurations that can be assigned to
its edges. This result is a necessary condition for applying many
statistical mechanical methods such as the Metropolis algorithm to
computing the expectation values of observables. Remarkably, this
result allows one in certain cases to extract expectation values of
observables depending on the original representation variables
(and possibly on the hyperboloid variables as well)~\cite{wade2}.

We note that the face factoring transformation to hyperboloid variables 
and the positivity results are valid for models with choices of edge and
face amplitudes different from the Perez-Rovelli model
(see Subsections~\ref{ss:2} and~\ref{ss:3}).  For example, a
simplified model with the edge amplitude trivialized to $\A_e=1$
and face amplitude taken to be $\A_f=1/\pf^2$ is being used to test 
numerical applications of positivity~\cite{wade2}. 

\subsection{Positivity for Manifolds with Boundary}
\label{ss:boundary}

We now turn to triangulations of 4-manifolds 
with boundary; in spin foam quantum gravity these are
interpreted as histories that interpolate between incoming and outgoing
boundary 3-geometries. 

The amplitudes for histories are generally required to satisfy
a composition law.  That is, if $\Delta_1$ is a triangulation
with initial boundary $\Gamma_1$ and final boundary $\Gamma_2$,
and $\Delta_2$ is a triangulation with initial boundary $\Gamma_2$
and final boundary $\Gamma_3$, then the amplitude for the 
combined triangulation 
$\Delta_1 \coprod_{\Gamma_2} \Delta_2$
is required to be the integral of $\A(\Delta_1) \A(\Delta_2)$
over all boundary data on $\Gamma_2$.
(The dependence on the boundary data has been suppressed from
the notation.)

In the original formulation in terms of representation
variables, triangles in the boundary tetrahedra are assigned 
fixed representation labels as boundary data.  
If this convention is still used, then one obtains an amplitude
$\A(\Delta, \{\pf\}_{f \in \partial F})$ that depends on the
labels on the faces in the boundary.
The composition law is satisfied as long as in
Equation~(\ref{eq:BCstatesum}) the edge amplitudes $\A_e$
are replaced with $\sqrt{\A_e}$, when $e$ is in the boundary.
(If the face amplitudes are non-trivial, they also must be
adjusted in the same way.)

Because of the factors $\sqrt{\A_e}$, this model cannot be converted
completely to face-factored form, as the integrals with respect to
$r_e$ cannot be brought outside of the square root.
However, it can be partially converted to face-factored form, with
some integrals with respect to $r_e$ variables left in place.
An additional difference between this form and
Equations~(\ref{eq:facefactoring}) and~(\ref{eq:Axei})
is that the integrations with respect to $\pf$ are omitted
for faces $f$ in the boundary.
Since the unintegrated face factors~(\ref{eq:facefactor}) are 
oscillatory functions, there is clearly interference between 
different $x_e^i$ configurations in this case.  
It is likely that after integrating out the
$x_e^i$ dependence there is no interference between the resulting 
amplitudes $\A(\Delta, \{\pf\}_{f \in \partial F})$, but our methods 
do not give any information about this.

However, since in most cases the majority of the variables lie in the interior
of the triangulation and hence give non-negative contributions, 
we are hopeful that statistical methods will be able to take advantage 
of this.  In any case, the fact that the integrated face-factors can
be analytically computed will be very useful for computations.

\subsubsection{A dual variables approach}
In the dual variables formulation of the Barrett-Crane model,
one regards the histories as being labelled by points in hyperbolic space.
From this point of view, it is really more natural to have
boundary data labelled by points in hyperbolic space as well,
and to integrate over the representation variables $\pf$
even for boundary faces.
With $\A_e = \Theta_4$, the natural boundary variables to hold
fixed in order to satisfy the composition law are the $r_e$ variables.
(It does not make sense to hold any of the $x_e^v$ variables fixed,
since they are not shared when two triangulations are joined along
a common boundary.)
However, the $\sqrt{\A_e}$ factors lead to complex-valued face-factors
which we are unable to analyze.

\longpage
If we simplify the model by setting $\A_e = 1$, then there are no
boundary variables.  The model can be converted fully to face-factored
form, and our methods show that the amplitudes are non-negative.
Incidentally, this implies that in the model with $\A_e = 1$ and
representation labels on the boundary, the integral of
$\A(\Delta, \{\pf\}_{f \in \partial F})$ over all
boundary data is non-negative.
Note that for this model to be finite, we need $\A_f = 1/\pf^2$.

\section{Fourier transforms}
\label{se:ft}

For $a > 0$,
let $\chi_a$ be the rectangle function of the interval $[-a,a]$,
defined by
\[
  \chi_a(k) \equiv 
    \begin{cases}
       1 & \text{for $k \in (-a,a)$}\\
       1/2 & \text{for $k = \pm a$}\\
       0 & \text{otherwise} \quad 
    \end{cases}
\]
and let $\sinc(t) \equiv \sin(t)/t$. 

Define the Fourier transform $\ft(f)$ of a function $f \in \lir$ by
\[
  \ft(f)(k) \equiv \intii f(t) \, e^{-ikt} \dt .
\]
This integral converges to a finite value for each $k$, and defines
a \emph{continuous} function of $k$~\cite[Thm.\ 7.5]{rudin}, which
is in $\ltr$.

If $f$ is in $\ltr$, then we define the Fourier transform of $f$
in a more indirect way.  The product $\chi_a f$ (essentially, $f$
restricted to the interval $[-a,a]$) is in $\lir$, and so we
can define $\ft(\chi_a f)$ as above.  Since $\chi_a f$ is also
in $\ltr$, it follows that $\ft(\chi_a f)$ is in $\ltr$,
and we define the Fourier transform of $f$ to be
\[
  \ft(f) \equiv \lim_{a\ra\infty} \ft(\chi_a f) ,
\]
with the limit taken in $\ltr$.  So $\ft(f)$ is only defined up
to equivalence in $\ltr$, i.e., only up to changes on a set of
measure zero.

The inverse Fourier transform of an $L^1$ function $f$ is
\begin{equation}\label{eq:invft}
  \ftinv(f)(k) \equiv \frac{1}{2 \pi} \intii f(t) \, e^{ikt} \dt
             = \frac{1}{2 \pi} \ft(f)(-k) .
\end{equation}
The definition is extended to $L^2$ functions by the same
limiting procedure use to define the forward Fourier transform.

The Fourier transform is an isomorphism from $\ltr$ to $\ltr$,
and the inverse Fourier transform is its inverse~\cite[Thm.\ 7.9]{rudin}.

\begin{example}
The rectangle function $\chi_a$ is in $\lir$, and it is elementary
to check that its Fourier transform is given by
\[
  \ft(\chi_a)(k) = 2 a \sinc(at) .
\]
The function $a \sinc(at)$ is in $\ltr$ but not $\lir$.
To compute its Fourier transform, we use (\ref{eq:invft})
to find
\[
  \ftinv(\pi \chi_a)(k) = \frac{\pi}{2 \pi} \ft(\chi_a)(-k)
                    = a \sinc(at) . 
\]
It follows that the Fourier transform of $a \sinc(at)$ is given by
\[
  \ft(a \sinc(at))(k) = \pi \chi_a(k) .
\]
We have shown this is true as functions in $\ltr$, i.e., almost
everywhere.  One can in fact show that, suitably interpreted,
this is true for all $k$.
\end{example}

We recall the following fact relating the Fourier transform and the
convolution.  See for example \cite[Thm.\ 7.2]{rudin} and 
\cite[p. 478]{godesses}.

\begin{lem}
If 
\begin{enumerate}
\item $f$ and $g$ are in $\lir$, or
\item $f$ and $g$ are tempered distributions and at least one has
      compact support,
\end{enumerate}
then 
$
  \ft(f * g) = \ft(f) \ft(g) 
$.
\end{lem}

We will make use of the following Corollary. 

\begin{cor}\label{cor:ftconv}
If $F$ and $G$ are tempered distributions whose Fourier transforms 
satisfy (1) or (2) above, then 
\[
  \ft(F G) = \frac{1}{2 \pi} \ft(F) * \ft(G) .
\]
\end{cor}

A tempered distribution is a distribution that is bounded
by a polynomial at infinity.  Any $L^p$ function, 
$1 \leq p \leq \infty$, defines a tempered distribution.

\begin{proof}
Let $f = \ft(F)$ and $g = \ft(G)$.  Then $\ft(f*g) = \ft(f) \ft(g)$ 
by the lemma.  By~(\ref{eq:invft}),
$\ftinv(f*g) = 2 \pi \ftinv(f) \ftinv(g)$
and so $\ftinv(\ft(F)*\ft(G)) = 2 \pi F G$.
Applying the Fourier transform gives the desired result.
\end{proof}

\section{Face Factor Positivity}
\label{se:pos}

In the following subsections, we show that the integrated
face factors are non-negative for various choices of face amplitude.
In this section, we introduce new variables $a_i$ and $t$---the $a_i$ are non-negative real numbers
that correspond to the hyperbolic distances $\phi_v^f$ and the non-negative
radial variables $r_e$.  The $p_f$ variable at a face is represented
in our proofs by $t$, symbolizing the ``time'' domain or original domain
to which Fourier transforms are applied. 

\subsection{Positivity for $A_F=1$.} \label{ss:1}

This choice of amplitude, which is that of the original Perez-Rovelli model,
leads to face factors of the form 
\[
  F_n^2(a_1, \ldots, a_n, t) = t^2 \prodin a_i \sinc(a_i t) .
\]
We wish to show that the following quantity is non-negative, 
for all choices of the $a_i \geq 0$:
\[
  I_n^2(a_1, \ldots, a_n) \equiv \intzi F_n^2(a_1, \ldots, a_n, t) \dt
  = \frac{1}{2} \intii F_n^2(a_1, \ldots, a_n, t) \dt .
\]
Here we have used that $F_n^2$ is an even function of $t$.

When $n=3$, it is straightforward to perform the integration and
show that the result is non-negative.
 
We now give a proof that $I_n^2$ is non-negative for $n \geq 4$. 
Ignoring the positive factor $1/2$, $I_n^2$ is equal to
the value of the Fourier transform 
\[
  \ft\left(t^2 \prodin a_i \sinc(a_i t)\right)(k)
\]
at $k=0$.  

Our method will be to evaluate the Fourier transform
for general $k$ in a way that lets us see that the value at $k=0$
is non-negative.  During the calculation we will take the
Fourier transform of $L^2$ functions, so
we must be careful about evaluation at a specific point (like $k=0$).
Since $n \geq 4$, the function $t^2 \prod a_i \sinc(a_i t)$ is in
$\lir$, and so its Fourier transform is continuous in 
$k$~\cite[Thm.\ 7.5]{rudin}.
Thus if we ensure that our indirect computation
of the Fourier transform leads to a function continuous in $k$,
then the value of that function at $k=0$ will be equal to
$I_n^2$.

Our first step is to use Corollary~\ref{cor:ftconv}
as well as the fact that the Fourier transform of $t g(t)$
is $i \ddk \ft(g)(k)$, when $\ft(g)$ is continuously differentiable.  
We obtain
\[
\begin{split}
  \ft\bigg(t^2 \prodin a_i \sinc(a_i t)\bigg)(k)
  &= i \ddk \ft\bigg(t \prodin a_i \sinc(a_i t)\bigg)(k) \\
  &= - \dddk \ft\bigg(\prodin a_i \sinc(a_i t)\bigg)(k) \\
  &= - \frac{\pi}{2^{n-1}} \dddk \convin \chi_{a_i}(k) .
\end{split}
\]
Corollary~\ref{cor:ftconv} applies since the $\chi_{a_i}$'s
are $L^1$ functions.
Also note that since $n \geq 4$, $\convin \chi_{a_i}(k)$ has a
continuous second derivative.  This justifies the first
two equalities, and means that evaluating the last expression
at $k=0$ makes sense.

By Appendix~\ref{ap:evenbumps}, $\convin \chi_{a_i}$
is an even bump, so it has a maximum at the origin.
Hence we have
\[
  - \dddk \convin \chi_{a_i} \geq 0 .
\]
So we have shown that $I_n^0$ is non-negative.

\subsection{Positivity for $\A_F=\frac{1}{\pf}$}
\label{ss:2}

This choice of amplitude leads to face factors of the form 
\[
  F_n^1(a_1, \ldots, a_n, t) = t \prodin a_i \sinc(a_i t) .
\]
We wish to show that the following quantity is non-negative, for all
choices of the $a_i \geq 0$:
\[
  I_n^1(a_1, \ldots, a_n) \equiv \intzi F_n^1(a_1, \ldots, a_n, t) \dt .
\]

For $n=2$, these integrals converge conditionally when $a_0 \neq a_1$.
In this case, it is straightforward to perform the integration and
show that the result is non-negative.
However, when $a_0=a_1$, the integral is divergent. 
Because the $n=2$ case does not arise for any non-degenerate triangulation, 
the $a_0=a_1$ divergence is not a problem. 

We now proceed to prove positivity for $n \geq 3$.
To apply Fourier methods, we express $I_n^1$ as an integral over the real
line, by multiplying by the Heaviside function
\[
  H(t) \equiv \begin{cases}
                1 & t \geq 0 \\
                0 & t < 0 \quad .
              \end{cases}
\]
So we wish to show that
\begin{equation}\label{eq:int}
  \intii H(t) \, t \prodin a_i \sinc(a_i t) \dt
\end{equation}
is non-negative.  
This is the value of the Fourier transform 
\[
  \ft\left(H(t)\,t \prodin a_i \sinc(a_i t)\right)(k)
\]
at $k=0$.  

As in the Subsection~\ref{ss:1},
our method will be to evaluate the Fourier transform
for general $k$ in a way that lets us see that the value at $k=0$
is non-negative.  
Since $n \geq 3$, the function $H(t) \, t \prod a_i \sinc(a_i t)$ is in
$\lir$, and so its Fourier transform is continuous in $k$.
We will ensure that our computation
of the Fourier transform leads to a function continuous in $k$,
so that the value of that function at $k=0$ will be equal to
$I_n^1$.

Proceeding as in Subsection~\ref{ss:1},
we obtain
\begin{equation} \label{eq:prodtoconv}
\begin{split}
  \ft\bigg(H(t)\,t \prodin a_i \sinc(a_i t)\bigg)(k)
  &= \frac{1}{2 \pi} \ft(H)(k) \conv \ft\bigg(t \prodin a_i \sinc(a_i t)\bigg)(k) \\
  &= \frac{1}{2 \pi} \ft(H)(k) \conv i \ddk\bigg(\ft\bigg(\prodin a_i \sinc(a_i t)\bigg)(k)\bigg) \\
  &=  \frac{1}{2^n} \bigg(\pi \delk - \pvik \bigg) \conv 
     i \ddk\bigg(\convin \chi_{a_i}(k)\bigg) .
\end{split}
\end{equation}
Some comments are in order.  
First, since $n \geq 3$, Lemma~\ref{lem:chiconv} shows that 
$\convin \chi_{a_i}$ is continuously differentiable, 
which justifies the second equality.
Second, the Heaviside function $H$ is not in $\lpr$ for any $p$, 
and so we must regard it as a tempered distribution.  
Its Fourier transform involves $\delk$ and $\pvk$.
Here $\delk$ is the usual delta distribution and $\pvk$ is the 
``principal value of $\frac{1}{k}$'' distribution (see
Appendix~\ref{ap:pv}).
Third, since $\ft(H)$ is tempered and 
$\ft\big(t \prodin a_i \sinc(a_i t)\big)$ has compact support,
Corollary~\ref{cor:ftconv} justifies the first equality
in~(\ref{eq:prodtoconv}).

Observe that the integral~(\ref{eq:int}) is real, and so
we can ignore the imaginary part of~(\ref{eq:prodtoconv}).
(One can also use the methods below to see directly that the
imaginary part is continuous in $k$ and vanishes at $k=0$.)
So we focus on the real part of the right-hand side.
Ignoring positive constants, this is
\begin{equation}\label{eq:realpart}
  \pvk \conv g'(k) \quad\text{where}\quad 
    g(k) = \convin \chi_{a_i}(k) .
\end{equation}
By Corollary~\ref{cor:chievenbump}, $g$ is an
even bump.  Moreover, by Lemma~\ref{lem:chiconv}, $g'$ exists 
and is itself piecewise $C^1$.
So by the results of Appendix~\ref{ap:pv}, the convolution
$\pvk \conv g'(k)$, which is a priori a distribution,
is in fact the continuous function
$\pvintii \frac{g'(k-\ell)}{\ell} \dl$.
Thus we can sensibly evaluate this at $k=0$, which gives
\[
  \pvintii \frac{g'(-\ell)}{\ell} \dl .
\]
But since $g$ is an even bump, $g'(-\ell)$ is non-negative
when $\ell > 0$ and non-positive when $\ell < 0$.
Thus the integrand is non-negative for all $\ell \neq 0$,
and so the result is non-negative.

\subsection{Positivity for $A_F=\frac{1}{\pf^2}$.}
\label{ss:3}

This choice of amplitude leads to face factors of the particularly
simple form 
\[
  F_n^0(a_1, \ldots, a_n, t) = \prodin a_i \sinc(a_i t) .
\]
The integrated face factors are
\[
  I_n^0(a_1, \ldots, a_n) \equiv \intzi F_n^0(a_1, \ldots, a_n, t) \dt 
  = \intzi \prodin a_i \sinc(a_i t) \dt .
\]
These are shown to be non-negative, for all $n \geq 1$ and all 
$a_i \geq 0$, in~\cite{remarkable}, using methods similar
to those used here.  
In addition, they give a closed form for $I_n^0(a_1, \ldots ,a_n)$
and relate it to the volume of certain polyhedra and their intersections. 
See also~\cite{expmath} for an introduction to this material.

\section{Conclusions}
\label{se:concl}

We have shown that in the path integral for the Lorentzian
Barrett-Crane model, with various normalizations, the amplitude
$\A(\Delta)$ for a triangulation $\Delta$ of a closed 4-manifold
is always non-negative.
This provides further evidence for the 
conjecture~\cite{BaezChristensen} that the amplitude $\A(\Delta, \pf)$
for a triangulation labelled by representations is always non-negative,
as is the case for the Riemannian model.
Our method is to use Cherrington's face factoring approach
to express the amplitude in terms of hyperboloid
variables $x_e^i$ instead of representation variables, and to
show that $\A(\Delta, x_e^i)$ is non-negative.

The above results hold for the original Perez-Rovelli model as well as
those with non-standard face normalizations $\A_f=1/\pf$ and $\A_f=1/\pf^2$.
The methods could be used for other models as well.

The non-negativity of the amplitudes $\A(\Delta,x_e^i)$ is highly significant 
from the point of view of numerical computations, as
it allows statistical methods such as the Metropolis algorithm
to be used for efficient computation of expectation values of
observables.
Moreover, exact expressions for $\A(\Delta,x_e^i)$ can be found~\cite{wade}.
Computational results will be presented in a forthcoming paper~\cite{wade2}.

In the case of a 4-manifold with boundary, we find
that there is no interference in the interior of the spin
foam, which may still be useful from a computational point of view.

\subsection*{Acknowledgements}

The authors would like to thank Igor Khavkine, Gord Sinnamon and 
Josh Willis for helpful conversations.

\appendix
\section{Convolutions of $\chi_a$'s}\label{ap:evenbumps}

Since the Fourier transform of $\sinc(at)$ involves the rectangle
function $\chi_a$ of the interval $[-a,a]$, and we want to understand
integrals of products of $\sinc$ functions, we will study convolutions
of the $\chi_a$ functions.
A key property of $\chi_a$ that we will use is that it
is an ``even bump''.

\begin{defn*}
An \emph{even bump} is a function $e:\R \ra \R$ such that
\begin{enumerate}
\item $e(x) = e(-x)$ for all $x$ ($e$ is an even function).
\item $e(x) \geq 0$ for all $x$.
\item $e$ has compact support.
\item\label{it:dec} $e(x) \geq e(y)$ for $0 \leq x \leq y$.
\end{enumerate}
\end{defn*}

It follows that $e$ is bounded above by $e(0)$ and is in $\lir$.

\begin{lem}\label{lem:evenbumpconv}
The convolution of an even bump with an even bump
is again an even bump.
\end{lem}

We thank Gord Sinnamon for the elegant proof of condition~(\ref{it:dec}).

\begin{proof}
Let $e$ and $f$ be even bumps, and define $g$ to be the convolution:
\[
  g(x) = (e*f)(x) = \intii e(y)f(x-y) \dy .
\]
This makes sense for every $x$, since $e$ and $f$ are bounded
and have compact support.

We shall establish each of the four properties for $g$ in turn.

\begin{enumerate}
\item Using that $e$ and $f$ are even, and then substituting
$y = -y'$, we have
\[
  g(x) = \intii e(y)f(x-y) \dy = \intii e(-y)f(-x+y) \dy
  = \intii e(y') f(-x-y') \dy' = g(-x) .
\]
\item Since $e$ and $f$ are non-negative,
their convolution will clearly be non-negative.
\item Since $e$ and $f$ have compact support, 
their convolution will also have compact support.
\item It suffices to show that for $x > 0$ and $s > 0$,
$g(x-s) \geq g(x+s)$.  Evaluating the difference, we obtain
\[
\begin{split}
  &g(x-s) - g(x+s) = (e*f)(x-s) - (e*f)(x+s) \\
  &= \intii e(y)    (f(x-s-y)-f(x+s-y)) \dy  \\
  &= \intii e(x-y') (f(y'-s)-f(y'+s))   \dy' \qquad (y' = x-y)  \\
  &= \intzi e(x-y') (f(y'-s)-f(y'+s)) \dy' + \intzi e(x+y') (f(-y'-s)-f(-y'+s)) \dy' \\
  &= \intzi (e(x-y')-e(x+y'))(f(y'-s)-f(y'+s)) \dy' .
\end{split}
\]
In the last integral, $x$, $y'$ and $s$ are all non-negative. 
Thus $|x-y'| \leq |x+y'|$ and $|y'-s| \leq |y'+s|$, so it follows from
properties (1) and (4) that $e(x-y') \geq e(x+y')$ and $f(y'-s) \geq f(y'+s)$.
Therefore the last integrand is a non-negative function.  So
$g$ is a non-increasing function on the positive axis.
\end{enumerate}
We have shown that $g$ is an even bump.
\end{proof}

\begin{cor} \label{cor:chievenbump}
An $n$-fold convolution $\chi_{a_1} * \cdots * \chi_{a_n}$ of
rectangle functions $\chi_{a_i}$ is an even bump.
\end{cor}

\begin{proof}
Since a single rectangle function $\chi_a$ is an even bump, 
this follows inductively from Lemma~\ref{lem:evenbumpconv}.
\end{proof}

We also need to make use of smoothness properties of the
$n$-fold convolution of $\chi_a$'s.

\begin{lem}\label{lem:fchiconv}
Let $f \in \lir$.  Then $f * \chi_a$ is a continuous function.
If $f$ is continuous, then $f * \chi_a$ is differentiable,
and $\ddk (f*\chi_a)(k)$ is the continuous function
$f(k+a)-f(k-a)$.
\end{lem}

If follows that if $f$ is $C^n$ 
(i.e., $f$ is $n$-times continuously differentiable)
then $f * \chi_a$ is $C^{n+1}$.

\begin{proof}
Suppose $f$ is in $\lir$.  Then
\[
  (f * \chi_a)(k) 
  = \intii f(\ell) \chi_a(k-\ell) \dl 
  = \int_{k-a}^{k+a} f(\ell) \dl ,
\]
which is continuous as a function of $k$.  If $f$ is continuous,
then 
\[
  \ddk (f * \chi_a)(k) 
  = \ddk \int_{k-a}^{k+a} f(\ell) \dl
  = f(k+a)-f(k-a) .
\]
\end{proof}

\begin{defn}
A function $f : \R \ra \R$ is \dfn{piecewise $C^1$} if
the real line can be partitioned into closed intervals $[s,t]$
such that $f'(k)$ exists on $(s,t)$, the one-sided derivatives
exist at the end-points, and all of these taken together form
a continuous function on the closed interval $[s,t]$.  
It follows that $f$ is continuous on $\R$.
\end{defn}

\begin{lem}\label{lem:chiconv}
Consider the $n$-fold convolution $g = \convin \chi_{a_i}$.
If $n = 2$, then $g$ is piecewise linear.
If $n \geq 3$, then $g$ is $(n-2)$-times continuously differentiable, 
and the $(n-2)$nd derivative is piecewise linear.
In particular, for $n \geq 3$, $g'$ is piecewise $C^1$ since it
is either $C^1$ or piecewise linear.
\end{lem}

\begin{proof}
If $n=2$, one can check directly that $g$ is piecewise linear.
The second part follows by induction from the first part and the previous lemma.
\end{proof}

\section{Principal value of $1/k$}\label{ap:pv}

The Fourier transform of the Heaviside function $H$
involves the ``principal value of $1/k$'' distribution
denoted $\pvk$.  This distribution is defined by
\[
  \langle \pvk, \phi \rangle \equiv \pvintii \frac{\phi(k)}{k} \dk
  \equiv \limeps \intieei \frac{\phi(k)}{k} \dk .
\]
This makes sense when $\phi$ has a continuous first derivative
and compact support~\cite[Example 1, page 439]{godesses}.

For the following, refer to the definition of
piecewise $C^1$ from Appendix~\ref{ap:evenbumps}.

\begin{thm}\label{thm:pvkfcts}
If $f$ is piecewise $C^1$ and has compact support,
then the distribution $\pvk * f(k)$ comes from the
function $\pvintii \frac{f(\ell-k)}{k} \dk$, and this
function is continuous.
\end{thm}

In our applications, $f$ is the derivative of a convolution of at
least 3 rectangle functions, so by Lemma~\ref{lem:chiconv},
it is piecewise $C^1$.

\begin{proof}
Since $\pvk$ is tempered and $f$ has compact support,
the convolution is well-defined.  It is the distribution
such that
\[
  \langle \pvk * f(k), \phi \rangle 
  \equiv \langle \pvk \times f(\ell), \phi(k+\ell) \rangle 
  \equiv \langle \pvk, \langle f(\ell), \phi(k+\ell) \rangle \rangle 
\]
for any test function $\phi$ (smooth, with compact support).
Note that $\langle f(\ell), \phi(k+\ell) \rangle$ is also
a test function as a function of $k$.
Expanding the right-hand side gives
\[
\begin{split}
  &\langle \pvk, \intii f(\ell) \phi(k+\ell) \dl \rangle\\
  &= \langle \pvk, \intii f(\ell-k) \phi(\ell) \dl \rangle\\
  &= \pvintii \frac{1}{k} \intii f(\ell-k) \phi(\ell) \dl \dk \\
  &= \limeps \intieei \intii \frac{f(\ell-k) \phi(\ell)}{k} \dl \dk \\
  &= \limeps \intii \intieei \frac{f(\ell-k) \phi(\ell)}{k} \dk \dl \\
  &= \limeps \intii \phi(\ell) \feps(\ell) \dl ,
\end{split}
\]
where $\feps(\ell) = \intieei f(\ell-k)/k \dk$.
The exchange of integration in the second-last equality is justified
because the integrand is continuous with compact support,
since a strip around $k = 0$ is excluded.

By Lemma~\ref{le:limfepsl}, $\limeps \feps(\ell)$ exists pointwise.
By Lemma~\ref{le:unifbnd}, the functions $\phi \feps$
are bounded by an $L^1$ function.
So by Lebesgue's dominated convergence theorem, we 
can exchange the limit and the integration in the last
displayed equation.
Thus
\[
  \langle \pvk * f(k), \phi \rangle 
   = \intii \limeps \feps(\ell) \phi(\ell) \dl
   = \intii \pvintii \frac{f(\ell-k)}{k} \dk \, \phi(\ell) \dl .
\]
So the distribution $\pvk * f(k)$ comes from the function
$\pvintii \frac{f(\ell-k)}{k} \dk = \limeps \feps(\ell)$.

It remains to show that this function is continuous.
Since each $\feps$ is continuous, it suffices to show
that the convergence to $\limeps \feps$ is uniform. 
The method used in the proof of Lemma~\ref{le:unifbnd}
shows that
\[
  |\limeps \feps(\ell) - F_\delta(\ell)| \leq 2 \delta \sup |f'| ,
\]
where $\sup |f'|$ is defined there as well.

We have shown that 
the distribution $\pvk * f(k)$ comes from the continuous function
$\pvintii \frac{f(\ell-k)}{k} \dk$.
\end{proof}

\begin{lem}\label{le:limfepsl}
  Let $f$ be a function with compact support such that both one-sided
  derivatives exist everywhere.  Then, using the notation above,
  $\limeps \feps(\ell)$ exists for each $\ell$.
\end{lem}

\begin{proof}
Fix $\ell$ and choose $A$ large enough so that the support
of $f(\ell-k)$ as a function of $k$ lies in $[-A,A]$.  
($A$ will depend on $\ell$.)
By symmetry,
\[
\intAeeA \frac{f(\ell)}{k} \dk = 0 .
\]
Therefore
\[
\begin{split}
\limeps \feps(\ell) 
&= \limeps \intieei \frac{f(\ell-k)}{k} \dk \\
&= \limeps \intAeeA \frac{f(\ell-k)}{k} \dk \\
&= \limeps \intAeeA \frac{f(\ell-k)-f(\ell)}{k} \dk.
\end{split}
\]
The last integrand is clearly continuous as a function of $k$ on $(0,A]$.
Since $f$ has one-sided derivatives everywhere, it extends to a 
continuous function on $[0,A]$.  Thus
\[
  \limeps \int_{\epsilon}^A \frac{f(\ell-k)-f(\ell)}{k} \dk
\]
exists.  Similarly, 
\[
  \limeps \int_{-A}^{\epsilon} \frac{f(\ell-k)-f(\ell)}{k} \dk
\]
exists.
\end{proof}

\begin{lem}\label{le:unifbnd}
Let $f$ be piecewise $C^1$ with compact support and let
$\phi$ be a continuous function with compact support.
Then, there is a constant $E$ such that for all $\epsilon > 0$,
$\phi(\ell) \feps(\ell) \leq E \phi(\ell)$.
In particular, the family of functions $\phi \feps$ is dominated
by an $L^1$ function.
(The notation $\feps$ is from the proof of Theorem~\ref{thm:pvkfcts}
and is recalled below.)
\end{lem}

\begin{proof}
Choose $B$ and $C$ so that the support of $f$ is contained in
$[-B,B]$ and the support of $\phi$ is contained in $[-C,C]$.
Let $A = B + C$.  Then for all $\epsilon > 0$ and all $\ell$ in $[-C,C]$,
\[
\begin{split}
\feps(\ell) 
&\equiv \intieei \frac{f(\ell-k)}{k} \dk \\
&= \intAeeA \frac{f(\ell-k)}{k} \dk \\
&= \intAeeA \frac{f(\ell-k)-f(\ell)}{k} \dk .
\end{split}
\]
Therefore, 
\[
\begin{split}
|\feps(\ell)| 
&\leq \intAeeA \left|\frac{f(\ell-k)-f(\ell)}{k}\right| \dk \\
&\leq \intAeeA \sup |f'| \dk \\
&\leq 2C \sup |f'| ,
\end{split}
\]
where the supremum is taken over $\R$ (or $[-C,C]$) and $|f'|$ means 
the maximum of the absolute values of the one-sided derivatives of $f$.
This supremum is finite, since $f$ is piecewise $C^1$ and has compact 
support.

Since this bounds $\feps(\ell)$ for $\ell$ in the support of $\phi$,
we get that $\phi(\ell) \feps(\ell) \leq 2 C \sup |f'| \phi(\ell)$
for all $\ell$.
\end{proof}

\end{document}